\documentclass{INTERSPEECH2023}
\usepackage{cite}
\usepackage{amssymb}
\usepackage{amsfonts}

\usepackage{multirow}
\usepackage{makecell}
\usepackage{color}
\usepackage{pifont}
\usepackage{comment}
\usepackage{soul}
\usepackage{amsmath,graphicx}
\usepackage{array} 
\interspeechcameraready 
\newcommand{\cmark}{\ding{51}}%
\newcommand{\xmark}{\ding{55}}%

\title{Towards Effective and Compact Contextual Representation\\ for Conformer Transducer Speech Recognition Systems}
\name{Mingyu Cui$^1$, Jiawen Kang$^1$, Jiajun Deng$^1$, Xi Yin$^3$, Yutao Xie$^3$, Xie Chen$^{2,3}$, Xunying Liu$^1$}
\address{
  $^1$The Chinese University of Hong Kong, Hong Kong SAR, China\\
  $^2$ MoE Key Lab of Artificial Intelligence, AI Institute,  \\ X-LANCE Lab, Department of Computer Science and Engineering, \\ Shanghai Jiao Tong University, Shanghai, China \\
  $^3$ Peng Cheng Laboratory, Shenzhen, China}
\email{}

\begin{document}

\bstctlcite{IEEEexample:BSTcontrol}
\ninept

\maketitle
% add nine font
\begin{abstract}
Current ASR systems are mainly trained and evaluated at the utterance level. Long range cross utterance context can be incorporated. A key task is to derive a suitable compact representation of the most relevant history contexts. In contrast to previous researches based on either LSTM-RNN encoded histories that attenuate the information from longer range contexts, or frame level concatenation of transformer context embeddings, in this paper compact low-dimensional cross utterance contextual features are learned in the Conformer-Transducer Encoder using specially designed attention pooling layers that are applied over efficiently cached preceding utterances’ history vectors. Experiments on the 1000-hr Gigaspeech corpus demonstrate that the proposed contextualized streaming Conformer-Transducers outperform the baseline using utterance internal context only with statistically significant WER reductions of 0.7\% to 0.5\% absolute (4.3\% to 3.1\% relative) on the dev and test data.

\end{abstract}


\begin{figure*}[htbp]
    \centering
    \vspace{-1mm}
    \setlength{\abovecaptionskip}{-0cm}
    \centering
    \includegraphics[width=7in]{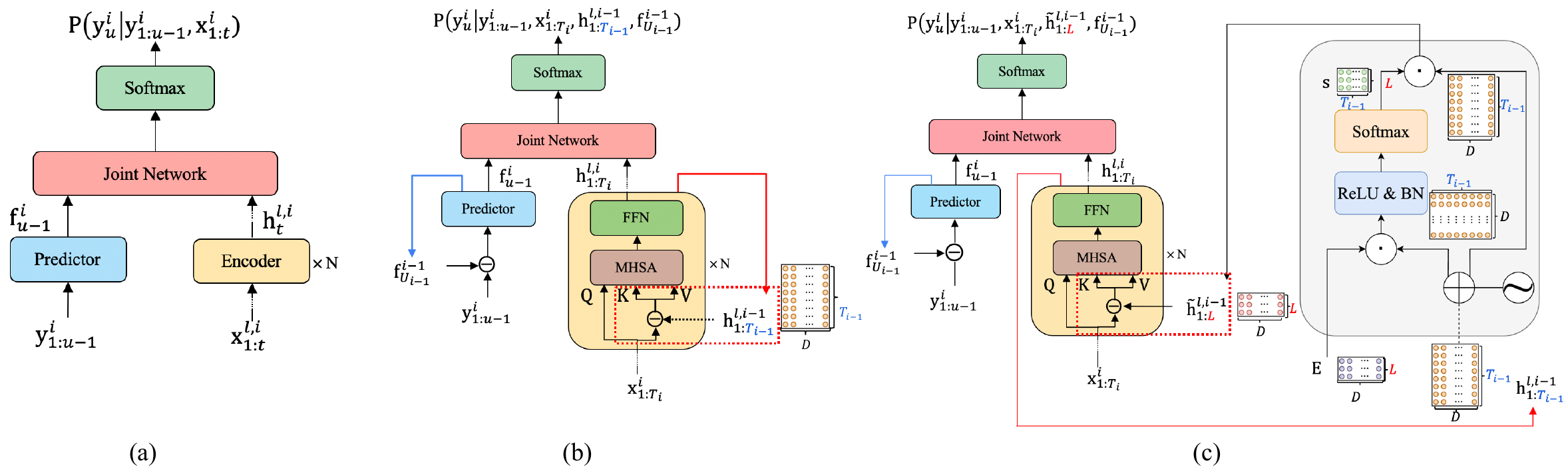}
    \caption{Examples of: a) Standard Conformer-Transducer models using utterance internal context only; (b) C-T models using cross utterance context of the most recent preceding $(i-1)^{\rm th}$ utterance $\mathbf{x}_{1:T_{i-1}}^{i-1}$ of $T_{i-1}$ frames in both the Encoder (connected via red line) and Predictor (via blue line encoding $(u-1)$ history words of current $i^{\rm th}$ utterance $\mathbf{x}^{i}_{1:T_{i}}$), where frame level concatenated Encoder contextual matrix ${\mathrm{h}^{l, i-1}_{1:T_{i-1}} \in ~\mathbb{R}^{T_{i-1} \times D}}$ are used; and (c) C-T models using the most recent preceding utterance's context in both Encoder and Predictor, where low-dimensional Encoder contextual representations $\mathrm{\Tilde{h}}^{l,i-1}_{1:L} \in \mathbb{R}^{L \times D}, L \ll T_{i-1}$ are compressed using attention pooling (circled in grey box, right) and $D$ is the Encoder output dimensionality. Grey dotted connections only used during forward passes in training. $\ominus$, $\odot$ and $\oplus$ denote matrix concatenation, multiplication and addition respectively. Encoder fusion of current and previous utterances' contexts is indicated in the red dotted box.
    %%means context encoding concatenation process.}
    }
    \vspace{-1mm}
    \label{fig:rnnt}
\vspace{-0.5cm}
\end{figure*}
\vspace{-0.2cm}
\section{Introduction}
\vspace{-0.2cm}
End-to-end (E2E) automatic speech recognition (ASR) technologies have achieved great success in recent years. A series of representative models, such as connectionist temporal classification (CTC) \cite{graves2012connectionist, watanabe2017hybrid}, listen-attend-spell (LAS) \cite{chan2016listen}, transformer \cite{vaswani2017attention,dong2018speech, karita2019comparative}, convolution-augmented transformer (Conformer) \cite{gulati2020conformer, guo2021recent} and recurrent neural network transducer (RNN-T) \cite{graves2012sequence, medsker2001recurrent, sak2014long, zhou2022efficient} have been developed. Among these, transformer based models, in particular those based on Conformer Encoder architectures \cite{tuske2021limit,zeineldeen2022improving, deng2022confidence, saon2023diagonal}, have demonstrated performance improvements over RNN based models.
%%and become the mainstream backbone ASR architectures.
 
It is well known that context plays an important role in human communication. A rich taxonomy of contextual cues across neighbouring speech utterances at acoustic-phonetic, prosodic, lexical, semantic and discourse level are used to determine what is likely to be said in a conversation. However, the majority of current ASR systems are trained and evaluated at the utterance level. To this end, the incorporation of long range, cross utterance contexts in E2E ASR systems provides a powerful solution. For this reason, contextual ASR models are attracting increasing research interest \cite{wei2022leveraging, kim2018dialog, chen2021developing, chang2021context, sun22_interspeech, hou2022bring,kim2019gated}. For example, the benefit from incorporating such cross-utterance information has been widely demonstrated on language modelling \cite{irie2019training,chen2020lstm, xiong2018session, dai2019transformer, kim2019end, liu2020contextualizing, liu2013use, beltagy2020longformer,sun2021transformer}. In contrast, limited prior researches in this direction have been conducted for Transformer or Conformer models \cite{rae2019compressive,fan2019speaker, tsunoo2019transformer, hori2020transformer, hori2021advanced}.

A key task in modelling cross utterance contexts for E2E ASR systems in general, including Transformers and Conformers, is to derive a suitable representation of the most relevant portion of history contexts to improve the prediction of current outputs, while incurring minimal computational overhead.  In LSTM-RNN based neural transducers \cite{hou2022bring}, the recurrent hidden vectors are used to encode preceding utterance histories, while attenuating the contribution from longer range contexts with an undesirable diminishing effect. 

Although the convolutional and attention mechanisms used in the Conformer architectures can capture both local and global feature patterns within a single utterance, there has no principled and well established solution when using these to model cross utterance contexts. A common practice \cite{hori2020transformer, hori2021advanced} is to utilize the outputs of Transformer or Conformer in each frame time step before being concatenated and serving as the long span context representation. However, there is a lack of mechanistic approaches to locate the most relevant portion of history contexts over time. In addition, this leads to computational efficiency and latency issues due to the high dimensionality of the frame level concatenated contextual representation that is dependent on the lengths of preceding utterances. %% being modelled. 

To this end, compact low-dimensional cross utterance contextual features are learned in the Encoder module of Conformer-Transducer (C-T) \cite{gulati2020conformer, zhang2020transformer} ASR systems in this paper using specially designed attention pooling layers applied over preceding utterances’ history vectors to auto-configure the context weighting at different time steps. Similar to the attentive pooling \cite{santos2016attentive} and attentive speaker embedding \cite{okabe2018attentive} used in speaker recognition \cite{campbell1997speaker} tasks, such attention based pooling compresses variable length hidden states into fixed length context representations. Inspired by Transformer-XL \cite{dai2019transformer} language models, the preceding utterances’ history vectors are efficiently cached prior to the pooling to improve computational efficiency. Cross utterance contexts are also incorporated into the Predictor. Experiments on the 1000-hr Gigaspeech corpus demonstrate that the proposed cross utterance context conditioned Conformer-Transducer system outperforms the baseline using utterance internal context only with statistically significant word error rate (WER) reductions of 0.7\% to 0.5\% absolute (4.3\% to 3.1\% relative) on the dev and test data, while incurring a moderate processing latency increase by 7.5\% during cross utterance context fusion.

The main contributions of this paper are summarized as follows. First, to the best of our knowledge, this is the first work to efficiently model attention pooling compressed low-dimensional cross utterance contexts in Conformer-Transducer systems. In contrast, related prior researches focused on other architectures based on, for example, Transformer and Conformer-transformer AED \cite{rae2019compressive, tsunoo2019transformer, hori2020transformer, hori2021advanced} models. Second, the compact low-dimensional cross utterance contextual features of in this paper enhances the computational efficiency and practical deployment of Conformer-Transducer and other  Transformer Encoder based E2E systems. Their benefit from cross utterance context modelling can be exploited to process streaming conversational data in naturalistic application scenarios. 
\vspace{-0.2cm}
\section{Conformer Transducer ASR Architecture}
\vspace{-0.15cm}
\subsection{Neural Transducer}
\vspace{-0.15cm}
In this paper, the neural transducer \cite{graves2012sequence} model is adopted for speech recognition. The neural transducer consists of three components, which are audio "Encoder", text "Predictor" and "Joint Network" modules respectively, as shown in Figure \ref{fig:rnnt}(a). Here we denote  $\mathbf{x}_{1:T_i}^{i}=[\mathbf{x}_{1}^{i}, \mathbf{x}_{2}^{i}..., \mathbf{x}_{T_i}^{i}]$ as the $i$-th utterance of an audio clip or conversation session with $T_i$-frames and $\mathbf{y}_{1:U_i}^{i}=[\mathbf{y}_{1}^{i}, \mathbf{y}_{2}^{i}..., \mathbf{y}_{U_i}^{i}]$ as the corresponding label of length $U_i$. The acoustic feature sequence $\mathbf{{x}}_{1:t}^{i}$ is fed into Encoder to produce the acoustic representation $\mathbf{{h}}_{t}^{i}$. The history output labels $\mathbf{{y}}_{1:u-1}^{i}$ are fed into the Predictor module to generate the text representation $\mathbf{f}_{u-1}^{i}$. The outputs of Encoder and Predictor are then combined in the Joint Network via a non-linear function such as ReLU to obtain the hidden state $\mathbf{g}_{t, u-1}^{i}$ at time step $t$ with output history $\mathbf{y}_{1:u-1}^{i}$. These operations are as follows,
\vspace{-0.15cm}
\begin{equation}
    \begin{aligned}
        \mathbf{h}_t^{i} &= \mathrm{Encoder}(\mathbf{x}_{1:t}^{i}) \\ 
        \mathbf{f}_{u-1}^{i} &= \mathrm{Predictor}(\mathbf{y}_{1:u-1}^{i}) \\
        \mathbf{g}_{t,u-1}^{i} &= \mathrm{relu}(\mathbf{h}_t^{i} + \mathbf{f}_{u-1}^{i}) \\
        P(\mathbf{y}_{u}^{i}| \mathbf{y}_{1:u-1}^{i}, \mathbf{x}_{1:t}^{i}) &= \mathrm{softmax}(\mathbf{W}_{o} * \mathbf{g}_{t,u-1}^{i})
    \end{aligned}
\vspace{-0.1cm}
\end{equation}
where $\mathbf{W}_{o}$ is a linear transformation applied prior to the final Softmax output layer. Among existing neural transducer systems, RNN or LSTM \cite{graves2012sequence, hou2022bring} and Transformer \cite{chen2021developing, zhang2020transformer, yeh2019transformer} architectures have been used for the Encoder, while the Predictor module is commonly based on LSTM. In this paper, Conformer-Transducers (C-T) designed using Conformer based Encoder and LSTM Predictor modules are used throughout this paper. An example Conformer-Transducer model using utterance internal context only is shown in Fig. \ref{fig:rnnt}(a).
\vspace{-0.15cm}
\subsection{Conformer Transducer}
\vspace{-0.15cm}
More specifically, the Conformer based Encoder is based on a multi-block stacked architecture. Each block contains the following components in turn: a position wise feed-forward (FFN) module, a multi-head self-attention (MHSA) module, a convolution (CONV) module and a final FFN module at the end. Among these, the CONV module consists of several modules: a 1-D pointwise convolution layer, a gated linear units (GLU) activation \cite{dauphin2017language}, a second 1-D point- wise convolution layer followed by a 1-D depth-wise convolution layer, a Swish activation and a final 1-D pointwise convolution layer. Layer normalization (LN) and residual connections are applied to stabilize the training and allow more stacked layers. For a given input feature sub-sequence $\mathbf{x}_{1:t}^{i}$ at time step $t$  fed into a Conformer Encoder, the vector output $\mathbf{h}_t^{l,i}$ of Conformer Encoder $l$-th block is:
\vspace{-0.15cm}
% \begin{equation}
%     \begin{aligned}
%         \hat{\mathbf{x}}_t &= \mathbf{x}_{1:t} + \frac{1}{2}\mathrm{FFN}(\mathbf{x}_{1:t}) \\
%         \mathbf{q}_t, \mathbf{k}_t, \mathbf{v}_t &= {\hat{\mathbf{x}}}_t\mathbf{W}_{q}, {\hat{\mathbf{x}}}_t\mathbf{W}_{k} , {\hat{\mathbf{x}}}_t\mathbf{W}_{v} \\
%         % \mathbf{x}^{'}_t &= \hat{\mathbf{x}} + \mathrm{MHSA}({\hat{\mathbf{x}}}W_q + {\hat{\mathbf{x}}}W_k + {\hat{\mathbf{x}}}W_v) \\
%         \mathbf{x}^{'}_t &= \hat{\mathbf{x}}_t + \mathrm{MHSA}(\mathbf{q}_t, \mathbf{k}_t, \mathbf{v}_t) \\
%         \mathbf{x}^{''}_t &= \mathbf{x}^{'}_t + \mathrm{Conv}(\mathbf{x}^{'}_t) \\
%         \mathbf{h}_t &= \mathrm{Layernorm}(\mathbf{x}^{''}_t + \frac{1}{2}\mathrm{FFN}(\mathbf{x}^{''}_t))
%     \end{aligned}
% \end{equation}
\begin{equation}
    \begin{aligned}
        {\hat{\mathbf{x}}}^{l, i, (0)}_{t} &= \mathbf{x}_{1:t}^{l-1,i} + \frac{1}{2}\mathrm{FFN}(\mathbf{x}_{1:t}^{l-1,i}) \\
        \mathbf{q}^{l,i}_{t}, \mathbf{k}^{l,i}_{ t}, \mathbf{v}^{l,i}_{t} &= {\hat{\mathbf{x}}}^{l, i, (0)}_{t}\mathbf{W}_{q}, {\hat{\mathbf{x}}}^{l, i, (0)}_{t}\mathbf{W}_{k} , {\hat{\mathbf{x}}}^{l, i, (0)}_{t}\mathbf{W}_{v} \\
        % \mathbf{x}^{'}_t &= \hat{\mathbf{x}} + \mathrm{MHSA}({\hat{\mathbf{x}}}W_q + {\hat{\mathbf{x}}}W_k + {\hat{\mathbf{x}}}W_v) \\
        \hat{\mathbf{x}}^{l, i, (1)}_t &= {\hat{\mathbf{x}}}^{l, i, (0)}_t + \mathrm{MHSA}(\mathbf{q}^{l,i}_{t}, \mathbf{k}^{l,i}_{t}, \mathbf{v}^{l,i}_{ t}) \\
        \hat{\mathbf{x}}^{l, i, (2)}_t &= \hat{\mathbf{x}}^{l, i, (1)}_t + \mathrm{Conv}(\hat{\mathbf{x}}^{l, i, (1)}_t) \\
        \mathbf{h}^{l,i}_{t} &= \mathrm{Layernorm}(\hat{\mathbf{x}}^{l, i, (2)}_t + \frac{1}{2}\mathrm{FFN}(\hat{\mathbf{x}}^{l, i, (2)}_t))
    \end{aligned}
\vspace{-0.2cm}
\end{equation}
where $\mathbf{W}_{q}$, $\mathbf{W}_{k}$ and $\mathbf{W}_{v}$ are the linear transformations to generate the query, key and value, respectively.\\

\begin{figure}[htbp]
    \centering
    \vspace{-1mm}
    \setlength{\abovecaptionskip}{-0cm}
    \centering
    \includegraphics[width=3.2in]{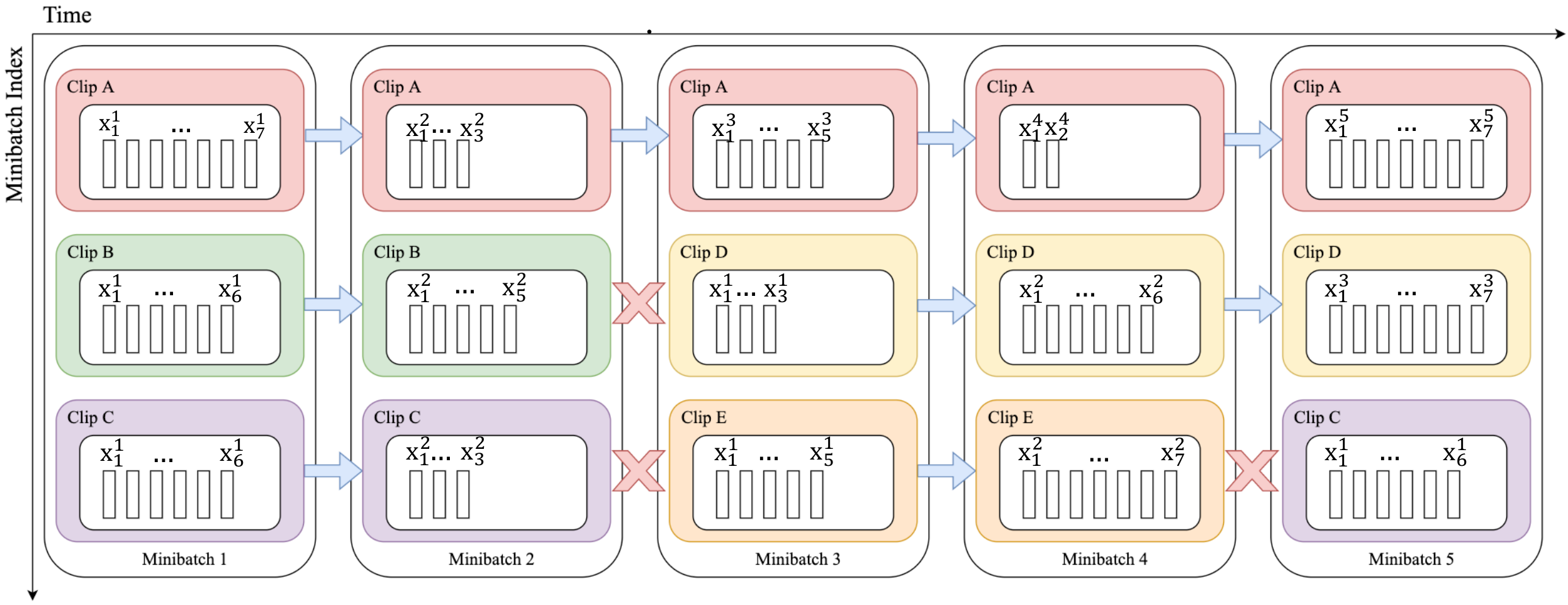}
    \caption{An example of data serialization for contextual C-T system training. The batch size is 3 and the utterances are from five audio clips A to E. Blue arrows indicate cross utterance contextual information between clip internal utterances is used, while red cross marks indicate it is not used when start processing the 1st utterance of a new clip.}
    \vspace{-0.5cm}
    \label{fig:data}
\end{figure}
\vspace{-0.1cm}
%% Conformer ASR Architecture}

\section{Compact Contextual Representation}
In this section, we propose cross utterance context conditioned Conformer-Transducer models with specially designed low-dimensional attention pooling layers to extract compact preceding utterances' Encoder contextual representations.

\subsection{Encoder Contextual Representation}
\vspace{-0.15cm}
Although the convolutional and attention mechanisms adopted in the Conformer architectures are used to capture both local and global feature patterns within a single utterance, there has no well-established solution when using them to model cross utterance contexts. A common practice \cite{dai2019transformer,hori2020transformer,hori2021advanced} is to utilize the outputs of Transformer or Conformer obtained at each frame of the preceding utterance(s). These are then being concatenated and serving as the long span context representation to augment the current utterance’s input features before applying the linear transformations to produce the query, key and value vectors.  In order to incorporate such frame level concatenated cross utterance Encoder contextual representations, the utterance internal context based C-T model of Eqn. (2) are now modified as 
\vspace{-0.2cm}
\begin{equation}
    \begin{aligned}
        {\hat{\mathbf{x}}}_{1:T_i}^{l,i} &= \mathbf{x}_{1:T_i}^{l-1,i} + \frac{1}{2}\mathrm{FFN}(\mathbf{x}_{ 1:T_i}^{l-1,i}) \\
        \hat{\mathbf{h}}^{l,i}_{1:T_i} &= \hat{\mathbf{x}}^{l,i}_{1:T_i} \ominus \mathrm{SG}(\mathbf{h}^{l,i-1}_{1:T_i}) \\
        \mathbf{q}^{l,i}, \mathbf{k}^{l,i}, \mathbf{v}^{l,i} &= {\hat{\mathbf{x}}}^{l,i}_{1:T_i}\mathbf{W}_{q}, {\hat{\mathbf{h}}^{l, i}_{1:T_i}}\mathbf{W}_{k} , {\hat{\mathbf{h}}^{l, i}_{1:T_i}}\mathbf{W}_{v} \\
    \end{aligned}
    \label{encoder}
\vspace{-0.2cm}
\end{equation}
%% where $\mathrm{SG}(\cdot)$ stands for stop-gradient, $\ominus$ indicates the concatenation of current input and previous hidden states along the length dimension.
where SG(·) stands for the “stop gradient” operator, $\ominus$ denotes matrix concatenation. An example of C-T models using such frame level concatenated preceding utterances’ Encoder contextual features are shown in Fig. 1(b). 
\vspace{-0.15cm}
\subsection{Compact Encoder Contextual Representation}
\vspace{-0.15cm}
The frame-level cross utterance Encoder contextual representations of Eqn.(3) above do not provide a mechanistic approach to locate the most relevant portion of preceding utterance contexts over time. Instead, the Encoder contextual representations obtained at all time steps of preceding utterances are retained non-discriminatively. The variable length nature of the resulting concatenated contextual vectors further leads to computational efficiency and scalability issues. 

In order to address the above issues, specially designed attention pooling layers are used in this paper and applied over preceding utterances’ Encoder contextual vectors to auto-configure the history context weighting at different time steps. The low-rank nature of these attention pooling layers allow variable length cross utterance Encoder contextual vectors used in Sec. 3.1 to be compressed to compact, low-dimensional features to condition the prediction of current utterance outputs for Conformer-Transducer systems. This design is inspired by the attentive pooling \cite{santos2016attentive} and attentive speaker embedding \cite{okabe2018attentive} used in speaker recognition \cite{campbell1997speaker} tasks. To further improve efficiency, akin to Transformer-XL \cite{dai2019transformer} language models, the preceding utterances’ Encoder hidden context vectors are efficiently cached prior to the attention based pooling operations. 
Let the C-T Encoder's outputs at the $l$-th Encoder layer be $\mathbf{h}^{l,i-1}_{1:{T_{i-1}}} \in \mathbb{R}^{T_{i-1} \times D}$ for the preceding $(i-1)^{\rm th}$ utterance of $T_{i-1}$ frames, where ${D}$ stands for the Encoder output vector dimensionality. The cross utterance Encoder contextual states are attention pooled and projected to low-dimensional representations $\mathbf{\Tilde{h}}^{l,i-1}_{1:L} \in \mathbb{R}^{L \times D}$ where and $L \ll T_{i-1}$, as 
%  \begin{equation}
%     \begin{aligned}
%         \mathbf{h}^{l,i-1}_{ 1:L} &= \mathbf{s} \cdot  {\hat{\mathbf{h}}}^{l, i-1}_{1:T_{i-1}}
%     \end{aligned}
% \vspace{-0.1cm}
% \end{equation}
% where (\textcolor{red}{merge Eqn. (4)-(5) into EnqArray})
\vspace{-0.2cm}
\begin{equation}
    \begin{aligned}
        \hat{\mathbf{h}}^{l, i-1}_{1:T_{i-1}} &= \mathrm{SG}(\mathbf{h}^{l, i-1}_{1:T_{i-1}}) \\
        \mathbf{\Tilde{h}}^{l,i-1}_{ 1:L} &= \mathrm{softmax}(\mathrm{bn}(\mathrm{relu}(\mathbf{E}(\hat{\mathbf{h}}^{l, i-1}_{1:T_{i-1}})^T))) \cdot  {\mathbf{h}}^{l, i-1}_{1:T_{i-1}} \\
    \end{aligned}
\vspace{-0.2cm}
\end{equation}
$\mathbf{E} \in \mathbb{R}^{L \times D}$ denotes the projection matrix to be learned, and $\mathrm{bn}(\cdot)$ stands for Batch Normalization. The resulting compact, fixed length $L \times D$ cross utterance Encoder contextual features are used together with the current utterance in C-T model training and evaluation. An example of C-T models using such compressed cross utterance Encoder contexts are shown in Fig. 1(c). 
\vspace{-0.35cm}
\subsection{Predictor Contextual Representation}
\vspace{-0.15cm}
In the LSTM based Predictor module, cross utterance contextual information can be further incorporated. A standard approach also considered in this paper is to cache the Predictor hidden vector state obtained at the end of the preceding utterance and concatenate it with the current Predictor input.  Given the input label sequence $\mathbf{y}_{1:u-1}^{i}$. The last Predictor hidden state computed from the previous utterance is cached as $\mathbf{f}^{i-1}_{U_{i-1}}$, before being concatenated with the current input $\mathbf{y}_{1:u-1}^{i}$ and fed into the Predictor module. The resulting cross utterance context conditioned Predictor vector outputs are
\vspace{-0.15cm}
\begin{equation}
    \mathbf{f}_{u-1}^i = \mathrm{Predictor}{(\mathbf{y}_{1:u-1}^{i}} \ominus \mathbf{f}^{i-1}_{U_{i-1}})
\vspace{-0.15cm}
\end{equation}
An example of using such preceding utterance’s context in a C-T system’s Predictor are in Fig. \ref{fig:rnnt}(b) and (c) (via blue lines).

\vspace{-0.15cm}
\subsection{Data Serialization}
\vspace{-0.15cm}

In traditional utterance-level based ASR training, we shuffle the data and construct multiple utterances in one minibatch based on their duration time. However, to capture cross utterance contexts,  we serialized the training data of the same audio clip or conversation session based on utterances’ start times. An example of data serialization for contextual C-T system training is shown in Fig. 2. The batch size is 3 and the utterances are from five audio clips A to E. Clip A contains five utterances. These 5 utterances in Clip A (in pink) are lined up based on their start times, while the cross utterance contexts are used as indicated by blue arrows. The red cross marks indicate cross utterance contexts are not used when starting processing the 1st utterance of a new clip, for example, E or D at minibatch 3. Since the number of utterances varies from the clips or sessions, short clips may not have enough utterances to fill the minibatches. In this case, utterances of other clips will be used to fill the minibatches and minimise synchronisation overhead.
\vspace{-0.2cm}
\section{Experiments}
\vspace{-0.1cm}
\subsection{Experiment Setup}
The Gigaspeech M size corpus \cite{chen2021gigaspeech} with 1000-hr speech collected from Audiobook, Podcast and YouTube is used for training. The dev and test sets randomly selected from Podcast and YouTube data containing 12 and 40 hours of speech were used.
In addition to the Conformer-Transducer system description of Section 2.1, raw speech were used as input features. The Conformer-Transducer model followed the ESPnet recipe \cite{guo2021recent} configuration. For C-T Encoder, we stacked 12 Encoder blocks where each Encoder block is configured with 8-head attention of 512-dim, and 2048 feed forward hidden nodes. For C-T Predictor, 1 uni-directional LSTM layer with 300 hidden size was adopted. 5000 byte-pair-encoding (BPE) tokens were served as the joint network outputs. The convolution subsampling module contains 2-D convolutional layers with kernel size 31. SpecAugment \cite{park2019specaugment} and dropout (rate set as 0.1) were used in training, together with model averaging performed over the last five epochs. Besides, we investigated streaming C-T model by applying a 1-frame self-attention look-ahead at each Encoder layer. 
%% We randomly sample 100 utterances from test set to measure run-time factors only during \textcolor{red}{the context encoding} forward process. 
% \textcolor{red}{Run time factors (RFTs) are measured on the test data for the computation incurred within the Encoder sub-layers where the previous and current utterance representations are fused (Fig. \ref{fig:rnnt} (b) and (c), red dotted boxes). }
Real time factors (RTFs) are measured on the test data for the computation incurred within the Encoder sub-layers where the previous and current utterance representations are fused (Fig. \ref{fig:rnnt} (b) and (c), red dotted boxes). 
All C-T systems with or without using cross utterance context are trained from scratch\footnote{Initializing cross utterance contextual C-T models with parameters of those without using context led to poor performance.}. The average utterance lengths are 3.96, 7.16 and 6.39 seconds for the training, Dev and Test sets respectively. Significant tests are performed using the standard NIST implemented \cite{pallet1990tools} Matched Pairs Sentence-Segment Word Error (MAPSSWE) Test proposed by Gillick \cite{gillick1989some} with a significant level of $\alpha=0.05$ denoted by $\dagger$ throughout the experiments . 
% All statistically significant WER reductions obtained over the baseline systems are marked using $\dagger$ in the following Tables.}

\vspace{-0.15cm}
\subsection{Evaluation Results}
\vspace{-0.4cm}
\begin{table}[htb]
\caption{WER\% \& RTFs of streaming Conformer-Transducers (C-T) without/with cross utterance context on the Gigaspeech M size corpus.}
%$\dagger$ denotes statistically significant WER differences (MAPSSWE, $\alpha= 0.05$ \cite{gillick1989some, pallet1990tools}) over the baseline (sys.1).}
\label{result}
\centering
\scalebox{0.61}{
\renewcommand\arraystretch{1.3}
\begin{tabular}{c|c|c|c|c|c|cc|c} 
\hline\hline
\multirow{3}{*}{ID} & 
\multirow{3}{*}{System} &
\multirow{2}{*}{Data} &
\multicolumn{2}{c|}{Encoder} &
Predictor  & 
\multicolumn{2}{c|}{WER\%} & 
\multirow{1}{*}{RTF} \\
\cline{4-8}
& &  \multirow{2}{*}{Repre.}& \#Prev.  & Context  & Prev. Utt. & \multirow{2}{*}{Dev} & \multirow{2}{*}{Test} & (Encoder\\
& & & Utt. & Vec. Len. & Context  &   & & Fusion)\\
\hline\hline

1 & &   & \multicolumn{3}{c|}{-} &  16.4 & 16.2 & 0.054 \\
%\cline{5-5} \cline{7-7} \cline{9-11}
\cline{4-9}
2 & \multirow{2}{*}{Streaming}&  & 1 & UttLen $\times$512  & \multirow{2}{*}{\cmark} &   $15.9^\dagger$ & $15.8^\dagger$ & 0.072  \\
%\cline{5-5} \cline{7-7} \cline{9-11}
3 & \multirow{2}{*}{C-T} & Clip  & 1 &\textbf{16} $\times$ 512&  &   16.2 & 16.0 & 0.058 \\
%\cline{5-5} \cline{7-7} \cline{9-11}
\cline{4-9}
4 &    &   & 2 & UttLen $\times$512 & \multirow{2}{*}{\cmark} & $\textbf{15.7}^\dagger$ & $\textbf{15.7}^\dagger$ & 0.076\\
%\cline{5-5} \cline{7-7} \cline{9-11}
5 &    &   & 2 & \textbf{32} $\times$ 512 &  & $15.9^\dagger$  & $15.8^\dagger$ & \textbf{0.058}\\
%\cline{5-5} \cline{7-7} \cline{9-11}
\hline\hline
\end{tabular}
}
\vspace{-0.2cm}
\end{table}
\begin{figure}[htbp]
    \centering
    % \vspace{-2mm}
    \setlength{\abovecaptionskip}{-0cm}
    \setlength{\belowcaptionskip}{-0cm}
    \centering
    \includegraphics[width=3in]{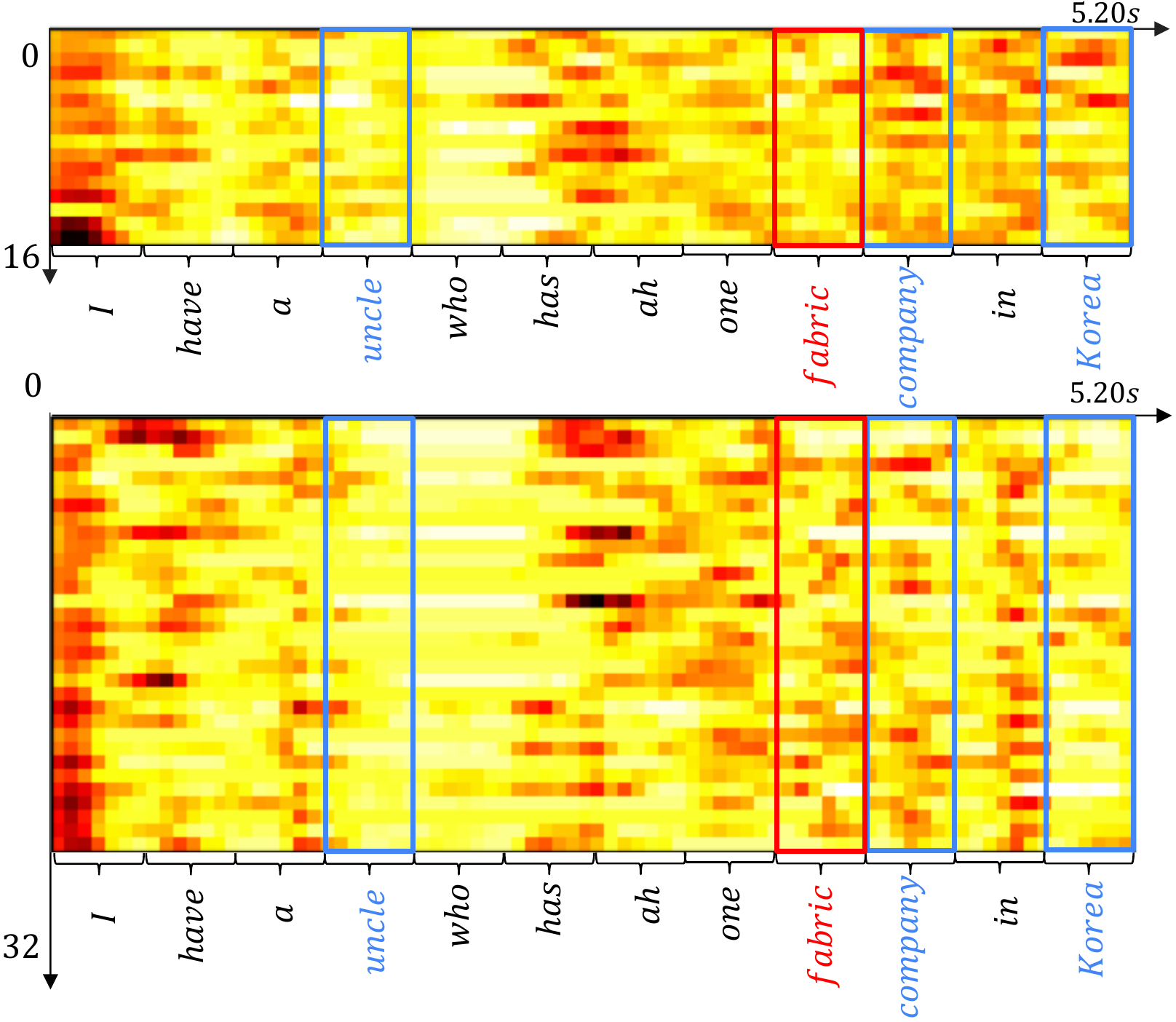}
    \caption{Examples of attention heat maps applied to the contextual representations (L=16, top; L=32, bottom) of a single previous utterance containing a sentence "I have a uncle who has ah one \textbf{fabric} company in Korea", which assign larger weights to words "\textbf{fabric}" in the contextual C-T model of Eqn. (4)-(5). The current utterance’s recognition outputs are "He always exports \textbf{fabric} to united states" and "He always exports \textbf{family} to united states" with or without using cross utterance context respectively.}
  %%  Attention heat maps from Sys.3 (Top) and Sys.5 (Bottom), where vertical axis is the compact context vector length equaling L=16 (Up), L = 32 (Bottom) respectively, horizontal axis is approximately time converting from embedding input.}
    \vspace{-0.3cm}
    \label{fig:heat}
\end{figure}

% baseline how : clip utt
% non streaming:
% trend1: only concat in (naive)encoder, decoder, or both -> better than baseline
% trend2: length=1, naive version better, compress version also better -> compress useful
% trend3: based on compress -> length=2, further better

% streaming:
% trend1: length 1: naive and compact both better than baseline
% trend2: length 2: further better

% no compress vs compress 345 vs 2
% compress pooling 678 vs 5 () performance add 4
% 91011 vs 678 degra, better, maybe equal 7 
% visual attention forward: 8 11 /
% streaming mask attention 
% significant test
% 13 vs 12 
% compress no compress
% compress no compress in length 2 but rtf 

% ct init 1, utterance internal different epoch
% from beginning
% move / foot note
% 
\vspace{-0.2cm}
The word error rates (WERs) and real-time factor (RTFs) of streaming C-T model without or with cross utterance contexts evaluated on Gigaspeech M size corpus are shown in Table \ref{result}. Several trends can be found. \textbf{1)} utilizing frame level concatenated cross utterance context states both in Encoder and Predictor modules (sys.2) outperforms the baseline C-T system without cross utterance context (sys.1) by statistically significant WER reductions of 0.5\% and 0.4\% absolute (3.0\% and 2.5\% relative) on the dev and test sets.  \textbf{2)} By using the attention pooled, 16$\times$512 or 32$\times$512-dimensional preceding utterance’s contextual features of Eqn. (4), similar performance improvements over the baseline C-T model using utterance internal context were obtained, in particular when the number of preceding utterances is increased to 2 (Sys. 5, 4 vs. 1).
%nature of the compressed cross utterance contextual features 
As expected, the compact method incurs a smaller increase in computation (measured in RTFs) during Encoder context fusion by 7.5\% over the baseline C-T, than that brought by frame level concatenation (Sys. 3, 5 vs. 2, 4).  
\textbf{3)} Increasing the number of preceding utterances produced consistent performances over only modelling the most recent one (Sys. 4, 5 vs. 2, 3), irrespective of whether frame level concatenated contextual features, or those that are attention pooled and compressed, are used. Finally, by visualizing the attention pooling heat maps of Sys. 3 and Sys. 5, interpretable weights assigned to an example preceding utterance’s word contents that intuitively leads to recognition accuracy improvements for a current utterance are shown in Fig. \ref{fig:heat}.
% fixing the history utterance length, we compact the context (sys.3) based on sys.2, which also outperforms the baseline system (sys.1) and obtains better processing latency of 13.0\% than the basic contextual system (sys.2). Third, based on sys.2 and sys.3 we enlarge the history utterance length from one to two (sys.4, 5), which outperform the baseline system consistently. Additionally, the best performance (sys.4) outperforms the baseline system (sys.12) by statistically significant WER reductions of 0.7\% and 0.5\% absolute (4.3\% and 3.1\% relative) on the dev and test sets. Additionally, by visualizing attention heat map of Sys.3 and Sys.5, we know that increasing the history utterance length and compacting it into fixed length enables the model to learn more widely context and attend to the important point of the original utterance. 
\begin{table}[htb]
\vspace{-0.35cm}
%% \caption{Performance (WER\% $\&$ RTF$s$ \textcolor{red}{in context encoding layer}) of non-streaming C-T model without / with contextual information evaluated on Gigaspeech M size corpus.}
\caption{WER\% \& RTFs of non-streaming C-T models w/o cross utterance context on the Gigaspeech M size corpus.}
%$\dagger$ denotes statistically significant WER differences (MAPSSWE, $\alpha= 0.05$ \cite{gillick1989some, pallet1990tools}) over the baseline (sys. 1).}
\label{nonstreaming}
\centering
\scalebox{0.61}{
\renewcommand\arraystretch{1.3}
\begin{tabular}{c|c|c|c|c|c|cc|c} 
\hline\hline
\multirow{3}{*}{ID} & 
\multirow{3}{*}{System} &
\multirow{2}{*}{Data} &
\multicolumn{2}{c|}{Encoder} &
Predictor  & 
\multicolumn{2}{c|}{WER\%} & 
\multirow{1}{*}{RTF} \\
\cline{4-8}
%% & &  \multirow{2}{*}{Repre.}& \#Prev.  & Context  & Prev. Utt. & \multirow{2}{*}{Dev} & \multirow{2}{*}{Test} & \\
%% & & & Utt. & Vec. Len. & Context &   & &\\
& &  \multirow{2}{*}{Repre.}& \#Prev.  & Context  & Prev. Utt. & \multirow{2}{*}{Dev} & \multirow{2}{*}{Test} &(Encoder\\
& & & Utt. & Vec. Len. & Context  &   & & Fusion)\\
\hline\hline
% 1 & \multirow{2}{*}{C-T}  & Utt. & &\multicolumn{2}{c|}{\multirow{2}{*}{-}} & \multicolumn{2}{c|}{\multirow{2}{*}{-}} & 14.3 & 14.2 & \\
6 &    & Utt. & \multicolumn{3}{c|}{\multirow{2}{*}{-}}   & 14.3 & 14.2 & -  \\
\cline{3-3}
% 2 &  & Clip&  &\multicolumn{2}{c|}{\multirow{2}{*}{-}} & \multicolumn{2}{c|}{\multirow{2}{*}{-}} & 14.2 & 14.0 &\\

7 &  &  \multirow{10}{*}{Clip}&  \multicolumn{3}{c|}{}&  14.2 & 14.0 & 0.058\\
\cline{4-9}
8 &  & & 1 & UttLen $\times$ 512 & \xmark&  14.0 & 13.9& 0.080\\
% \cline{5-5} \cline{7-7} \cline{9-10}
\cline{6-6} \cline{5-5}
9 &  &  & 0 & - & \multirow{2}{*}{\cmark} &   14.0 & 14.0 & 0.080\\
%\cline{5-5} \cline{7-7} \cline{9-11}
\cline{5-5}
10 &  Non-&  & 1 & UttLen $\times$ 512 &  & $14.0^\dagger$ & $13.8^\dagger$ & 0.080\\
\cline{4-9}

11 &   Streaming&  & 1 & \textbf{8} $\times$ 512 & \multirow{3}{*}{\cmark}& 14.1 & 13.9 & 0.062 \\
%\cline{5-5} \cline{7-7} \cline{9-11}
12 &  C-T &   & 1 & \textbf{16} $\times$ 512 &  & $14.0^\dagger$ & $13.8^\dagger$ & 0.064\\
%\cline{5-5} \cline{7-7} \cline{9-11}
13 &   &   & 1 & \textbf{32} $\times$ 512 &  &  14.1 & 13.9 & 0.067  \\
%\cline{5-5} \cline{7-7} \cline{9-11}
\cline{4-9}
14 &   &  & 2 & \textbf{8} $\times$ 512 & \multirow{3}{*}{\cmark} &  $14.0^\dagger$ & $13.8^\dagger$ & 0.062\\
%\cline{5-5} \cline{7-7} \cline{9-11}

15 &   &   & 2 & \textbf{16} $\times$ 512 &  &   14.1 & 13.9 & 0.064 \\
%\cline{5-5} \cline{7-7} \cline{9-11}
16 &   &   & 2 &\textbf{32} $\times$ 512&  &  $\textbf{14.0}^\dagger$ & $ \textbf{13.7}^\dagger$ & 0.067 \\
%\cline{5-5} \cline{7-7} \cline{9-11}
\hline\hline
\end{tabular}}
\vspace{-0.45cm}
\end{table}

Similar trends are observed on the experiments conducted using non-streaming C-T systems in Table \ref{nonstreaming}.  \textbf{1)} Utilizing cross utterance context the in Encoder (Sys. 8), Predictor (Sys. 9), or both (Sys. 10) consistently outperformed the baseline C-T without using such information (Sys. 6, 7).  \textbf{2)} Compressing the cross utterance contextual features via attention pooling produced comparable performance (Sys. 12 vs. 10), and again a smaller increase in RTF. \textbf{3)} Further increasing the number of history utterances being considered from one to two (Sys.14 - 16), the largest performance improvements over the baseline C-T systems (Sys. 16 vs. Sys. 6, 7) by statistically significant WER reductions of 0.3\% and 0.5\% absolute (2.0\% and 3.5\% relative) were obtained on the dev and test sets, respectively.

% baseline first
% 1. each better/ for example (return table1) 0.2 best
% 2. add lhuc trend not change/ better than baseline
% 3. cross adaptation better than baseline not better than multipass ->

\vspace{-0.25cm}
\section{Conclusions}
\vspace{-0.1cm}
In this paper, compact cross utterance contextual representations were incorporated into Conformer-Transducer (C-T) ASR systems using contextual attention pooling layers integrated with the C-T Encoder. Cross utterance contexts are also incorporated into the Predictor. Experiments on the 1000-hr Gigaspeech M corpus demonstrate that the proposed cross utterance context conditioned streaming Conformer-Transducer system outperform the baseline using utterance internal context only with statistically significant WER reductions of 0.7\% to 0.5\% absolute (4.3\% to 3.1\% relative) on the dev and test data, while incurring moderate increase of latency by 7.5\% in cross utterance context fusion. Future work will improve cross utterance contextual C-T models' generalisation and efficiency.

% In this paper, we presented the first use of system multi-pass rescoring and cross adaptation based on TDNN and Conformer ASR systems. Both of multi-pass rescoring and cross adaptation can utilize complementary but different information from each system which improve the performance. Experiments on 300-hr Switchboard corpus showed that the efficacy of the system combination methods and obtained 28.9\% relative WER reduction over the baseline Conformer system . Future research will focus on multi-pass rescoring more than two sub-systems.
\vspace{-0.2cm}
\section{Acknowledgements}
This research was supported by Hong Kong RGC GRF grant No. 14200021, 14200218, 14200220, Innovation \& Technology Fund grant No. ITS/218/21,  the National Natural Science Foundation of China  (No. 62206171), the International Cooperation Project of PCL, and Alibaba Group through Alibaba Innovative Research Program.
% \vspace{-0.1cm}

\bibliographystyle{IEEEtran}

\bibliography{mybib}

% \begin{thebibliography}{9}
% \bibitem[1]{Davis80-COP}
%   S.\ B.\ Davis and P.\ Mermelstein,
%   ``Comparison of parametric representation for monosyllabic word recognition in continuously spoken sentences,''
%   \textit{IEEE Transactions on Acoustics, Speech and Signal Processing}, vol.~28, no.~4, pp.~357--366, 1980.
% \bibitem[2]{Rabiner89-ATO}
%   L.\ R.\ Rabiner,
%   ``A tutorial on hidden Markov models and selected applications in speech recognition,''
%   \textit{Proceedings of the IEEE}, vol.~77, no.~2, pp.~257-286, 1989.
% \bibitem[3]{Hastie09-TEO}
%   T.\ Hastie, R.\ Tibshirani, and J.\ Friedman,
%   \textit{The Elements of Statistical Learning -- Data Mining, Inference, and Prediction}.
%   New York: Springer, 2009.
% \bibitem[4]{YourName17-XXX}
%   F.\ Lastname1, F.\ Lastname2, and F.\ Lastname3,
%   ``Title of your INTERSPEECH 2022 publication,''
%   in \textit{Interspeech 2022 -- 23\textsuperscript{rd} Annual Conference of the International Speech Communication Association, September 18-22, Incheon, Korea, Proceedings, Proceedings}, 2022, pp.~100--104.
% \end{thebibliography}

\end{document}